\documentclass[toc]{PoS}
\usepackage{wasysym}   

\title{Study of $\Sigma^+\pi^-$ Invariant Mass spectrum with the KLOE detector; preliminary results and possible hints for $\Sigma^+n$ internal conversion.}

\ShortTitle{Study of $\Sigma^+\pi^-$ Invariant Mass spectrum with the KLOE detector.}

\author{\speaker{A. Scordo$^a$}\\
C. Curceanu$^a$,K. Piscicchia$^{ab}$,I. Tucakovic$^a$,O. Vazquez Doce$^c$\\
on behalf of AMADEUS collaboration\\
\newline
{$^a$}INFN, Laboratori Nazionali di Frascati, Frascati (Roma), Italy\\
{$^b$}Centro Fermi - Museo Storico della Fisica e Centro Studi e Ricerche "Enrico Fermi", Piazza del Viminale 1 - Rome - 00184 - Italy\\
{$^c$}Excellence Cluster Universe, Technische Universit\"at M\"unchen, Garching,Germany\\
E-mail: \email{alessandro.scordo@lnf.infn.it}}

\abstract{The AMADEUS collaboration has the goal to perform unprecedented measurements in the field of the low-energy charged kaons-nuclei interactions, by implementing the existing KLOE detector with a dedicated setup in the inner region. As a preliminary step towards the realization, the AMADEUS team has analyzed the existent 2002-2005 KLOE data, studying the processes resulting from the negative kaons nuclear absorption  in the entrance wall of the KLOE Drift Chamber (containing mostly carbon) and in the gas filling it, mostly helium.  Processes containing $\Lambda p$ and $\Lambda d$ in the final state were looked for, together with the search for the $\Lambda(1405)$ going in both the neutral and the charged $\Sigma\pi$ channels. These analyses produced unique results, proving the possibility to obtain, for the first time, invariant mass spectra of the $\Lambda(1405)$ for all the possible decay channels. This was possible thanks to the unique features of the KLOE detector, including the excellent photon detection of its calorimeter. In addition to these results, other interesting effects like the $\Sigma-\Lambda$ internal conversion could be investigated. Preliminary results on the $\Sigma^+\pi^-$ decay channel and on the $\Sigma-\Lambda$ internal conversion will be presented.}

\FullConference{International Winter Meeting on Nuclear Physics,\\
		21-25 January 2013\\
		Bormio, Italy }

\begin{document}

\section{Introduction}

The KLOE \cite{KLOE} detector at DA$\Phi$NE \cite{DAF} represents a unique opportunity to perform a complete study of the $\Lambda(1405)$ resonance through all its three $\Sigma\pi$ decay channels simultaneously \cite{THEO1,THEO2,CLAS}. The importance of these measurements rely on the possibility to compare the different results for the various decay channels, investigating their contributions to the final resonance shape. The situation is complicated due to the presence of a second resonance, the $\Sigma(1385)$, whose contribution to the final $\Lambda(1405)$ spectra can not be easily substracted. 
Very promising and interesting results have been already obtained for the $\Sigma^0\pi^0$ decay channel \cite{KRIS}, to which, for isospin selection rules, the $\Sigma(1385)$ resonance can not decay. This work is focused on the preliminary analysis of the $\Sigma^+\pi^-$ decay channel for events in which a $K^-$ interacts with a proton in a $^{12}C$ nucleus; in this case, one has to take also into account the presence of the second resonance which affects the $\Sigma^+\pi^-$ invariant mass spectrum. Investigating all the possible background sources, some evidence of possible internal $\Sigma\rightarrow\Lambda$ conversion \cite{Internal} was found. In section 2, an overview of the KLOE detector is given, while in the third one the events selection is presented. In section four, the results of the MonteCarlo simulations and of the study of all possible background sources are discussed; in section five the internal conversion hypothesis and its consequences are explored and the resulting spectra are shown and discussed. 
In section 6, the final $\Sigma^+\pi^-$ invariant mass and momentum spectra are shown.

\section{DA$\Phi$NE collider and the KLOE detector}

\begin{figure}[htbp]
\centering
\mbox{\includegraphics[width=6.cm]{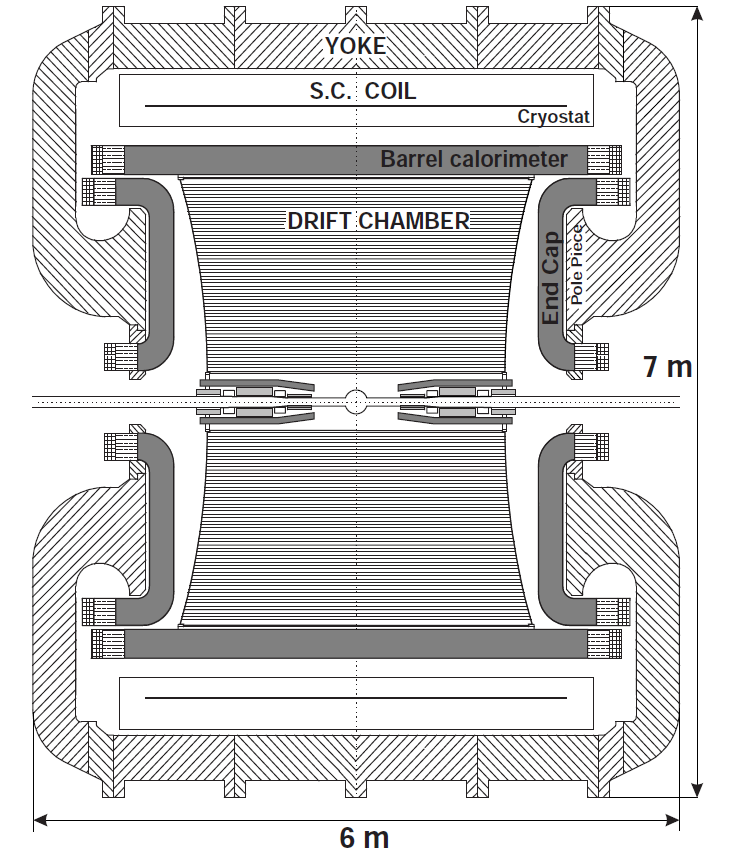}}
\caption{Section of the KLOE detector along the x-z plane.}
\label{KLOEdet}
\end{figure}

The KLOE detector is installed at the DA$\Phi$NE collider of the Laboratori Nazionali di Frascati-INFN \cite{DAF}, 
where the $\Phi$ resonance is created, almost at rest, via the collision between 510 MeV/c momentum electrons and positrons.
The decay products of the $\Phi$ resonance, mostly kaons and the particles reusulting from their interactions, are then detected by the KLOE detector.
The KLOE detector has a $\sim 4\pi$ geometry and an acceptance of 98\%; it consists of a large cylindrical Drift Chamber (DC) and a fine sampling
lead-scintillating fibers calorimeter, immersed in an axial magnetic field of 0.52 T provided by a superconducting solenoid, see fig. \ref{KLOEdet}

The DC \cite{DriftChamber} has an inner radius of 0.25m, an outer radius of 2m, a length of 3.3m
and is centered in the interaction point. The whole DC mechanics is made
of carbon fiber. The chamber is uniformly filled with 12582 drift cells, organized in coaxial
layers (12 inner and 46 outer). The gas mixture selected to fill the chamber is composed 90\% in volume of
$^4He$ and 10\% of $C_4H_{10}$ (Isobutane).
The chamber is characterized by excellent position and momentum resolutions. 
Tracks are reconstructed with a resolution in the transverse $R-\phi$ plane
$\sigma_{R\phi}\sim200\,\mu m$ and a resolution along the z-axis $\sigma_z\sim2\,mm$.
The transverse momentum resolution for low momentum tracks ($(50<p<300) MeV/c$)
is $\frac{\sigma_{p_T}}{p_T}\sim0.4\%$.

The KLOE calorimeter \cite{Calo} is composed of a cylindrical barrel and two endcaps,
providing a solid angle coverage of 98\%.
The volume ratio (lead/fibers/glue=42:48:10) is optimized for
a high light yield and a high efficiency for photons in the range
(20-300)MeV/c. The average density is 5 $g/cm^3$ and the radiation length is $X_0 \sim 1.5\,cm$.

\section{$\Sigma^+\pi^-$ events selection}

The main subject of this work is the study of the $\Lambda(1405)$ resonance through the $\Sigma^+\pi^-$ decay channel: 
$$ K^-p \rightarrow \Lambda(1405) \rightarrow \Sigma^+\pi^- \rightarrow (p\pi^0)\pi^-$$
where the $\pi^0$ is reconstructed trough the identification of the two photons ($\pi^0 \rightarrow \gamma\gamma$).
The analyzed data were taken from 2002 to 2005 by the KLOE collaboration and correspond to $\sim 560 pb^{-1}$.
The analysis is performed, in a first stage, only for events occuring in the KLOE Drift Chamber entrance wall which is mainly composed by $^{12}C$. 
A first event selection is performed requiring the presence of a $K^+$, which is always produced in pair with a $K^-$ from the decay of the $\Phi$ resonance \cite{PDG}; 
a further request is that the $K^-$ two body decay products are not detected and reconstructed, since that would be a clear signature of a $K^-$ not interacting with a proton.
Then we require the presence of at least one proton and one $\pi^-$; since the decay path of the $\Sigma^+$ is very short ($c\tau\sim2.4\,cm$ \cite{PDG}),
the proton and the pion are assumed to form a vertex in the KLOE events reconstruction algorithm.

\subsection{p - $\pi^-$ identification and $\Lambda(1116)$ rejection}

\begin{figure}[htbp]
\centering
\mbox{\includegraphics[width=7.5cm]{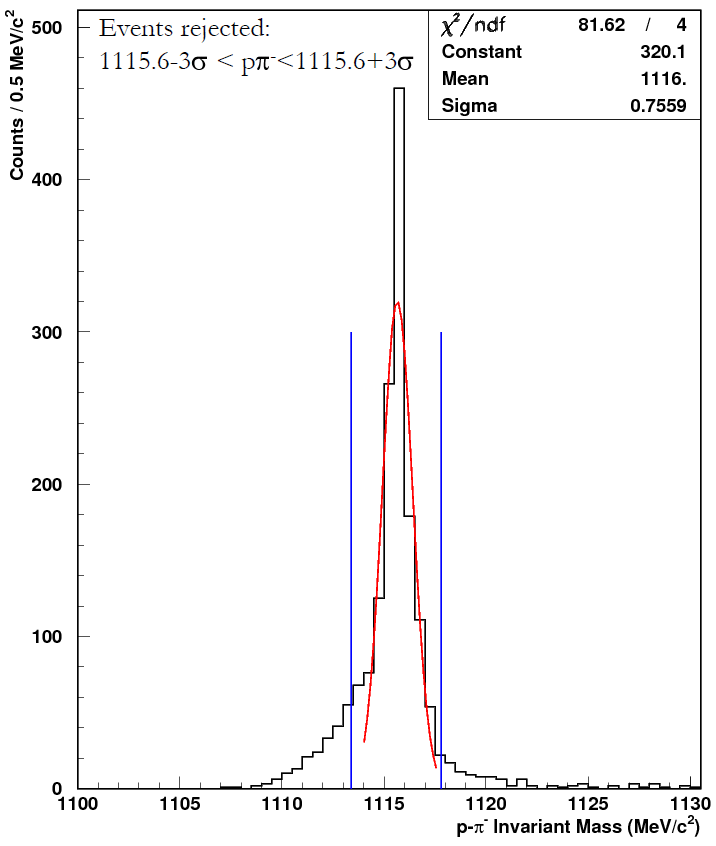}}
\caption{$p\pi^-$ invariant mass spectrum in MC simulated events of $\Lambda(1116)$. The mass range in which events are rejected is deduced from the fit parameters.}
\label{LAMBDAS}
\end{figure}

Protons and negative pions are identified from a two dimensional plot of their energy and momentum. 
The particle identification is done using energy information coming from the Drift Chamber wires and, only for the protons,
also from the Calorimeter, while the momenta are reconstructed from the tracks collected in the DC; the final selection is performed
via a two dimensional plot of energy and momentum.
Since the most important part of the background is represented by events involving the presence of a $\Lambda(1116)$, 
all the events in which the $p\pi^-$ invariant mass is compatible with a $\Lambda$ have to be rejected.
In order to do so, such events have been simulated in the KLOE detector and the resulting $p\pi^-$ invariant mass spectrum is shown in fig. \ref{LAMBDAS}

The spectrum is then fitted with a gaussian, determining the mass region in which events have to be excluded from the analysis in order to reject the $\Lambda(1116)$.

\subsection{Reconstruction of the $\pi^0$}

\begin{figure}[htbp]
\centering
\mbox{\includegraphics[width=15.cm]{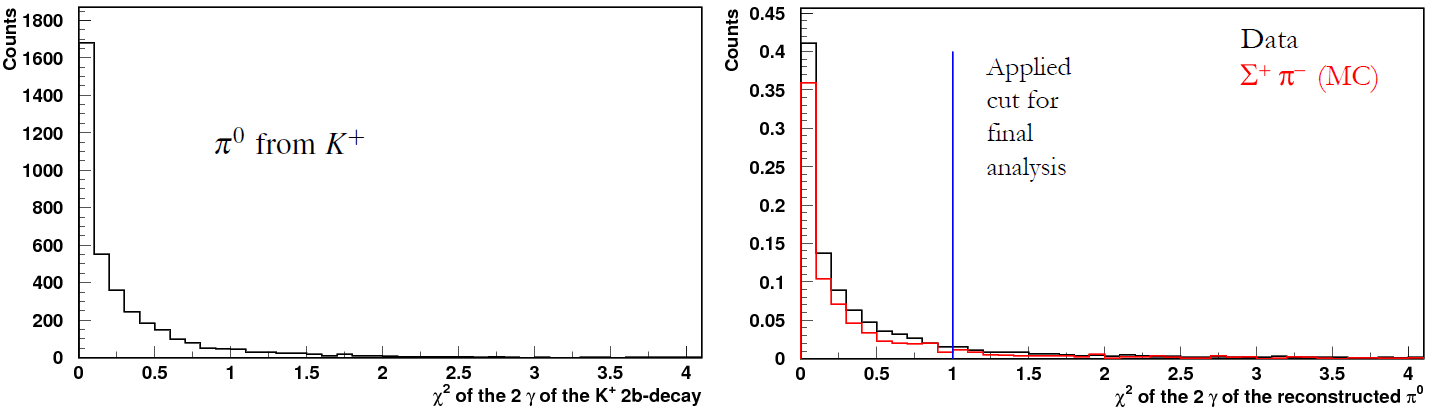}}
\caption{$\pi^0$ from $K^+$ decay (up) and $\pi^0$ of our interest (down) minimization variable. In the lower plot, $\Sigma^+\pi^-$ MC simulated $\pi^0$ spectrum (red) is superimposed to the real data one (black).}
\label{PI0}
\end{figure}

The $\pi^0$ are reconstructed using the KLOE calorimeter, searching for pairs of neutral clusters (without associated track in the D.C.) in time with the p$\pi^-$ vertex.
The selection of the best two clusters is done via the minimization, for each couple, of the quantity:

\begin{equation}
\chi^2=\frac{(t_{\gamma1}-t_{\gamma2})^2}{(\sigma_{t_{cl1}}+\sigma_{t_{cl2}})^2}
\end{equation}

\noindent where $t_{\gamma1,2}$ are the time in which the two photons cover the distance between the $p\pi^-$ vertex and the calorimeter,
with their propagated errors $\sigma_{t_{cl1,2}}$. 
In order to chose the best value for a correct selection/rejection of the reconstructed pion, the $\pi^0$ coming from the two-body decay of the $K^+$, 
identified and reconstructed by KLOE, can be used.
The resulting spectra are shown in fig. \ref{PI0}

According to the upper plot of fig.\ref{PI0}, a selection cut can be applied rejecting from the analysis all the events with $\chi^2\ge1$.

\section{Invariant Mass spectrum of $\Sigma^+ \rightarrow p \pi^0$}

\begin{figure}[htbp]
\centering
\mbox{\includegraphics[width=12.4cm]{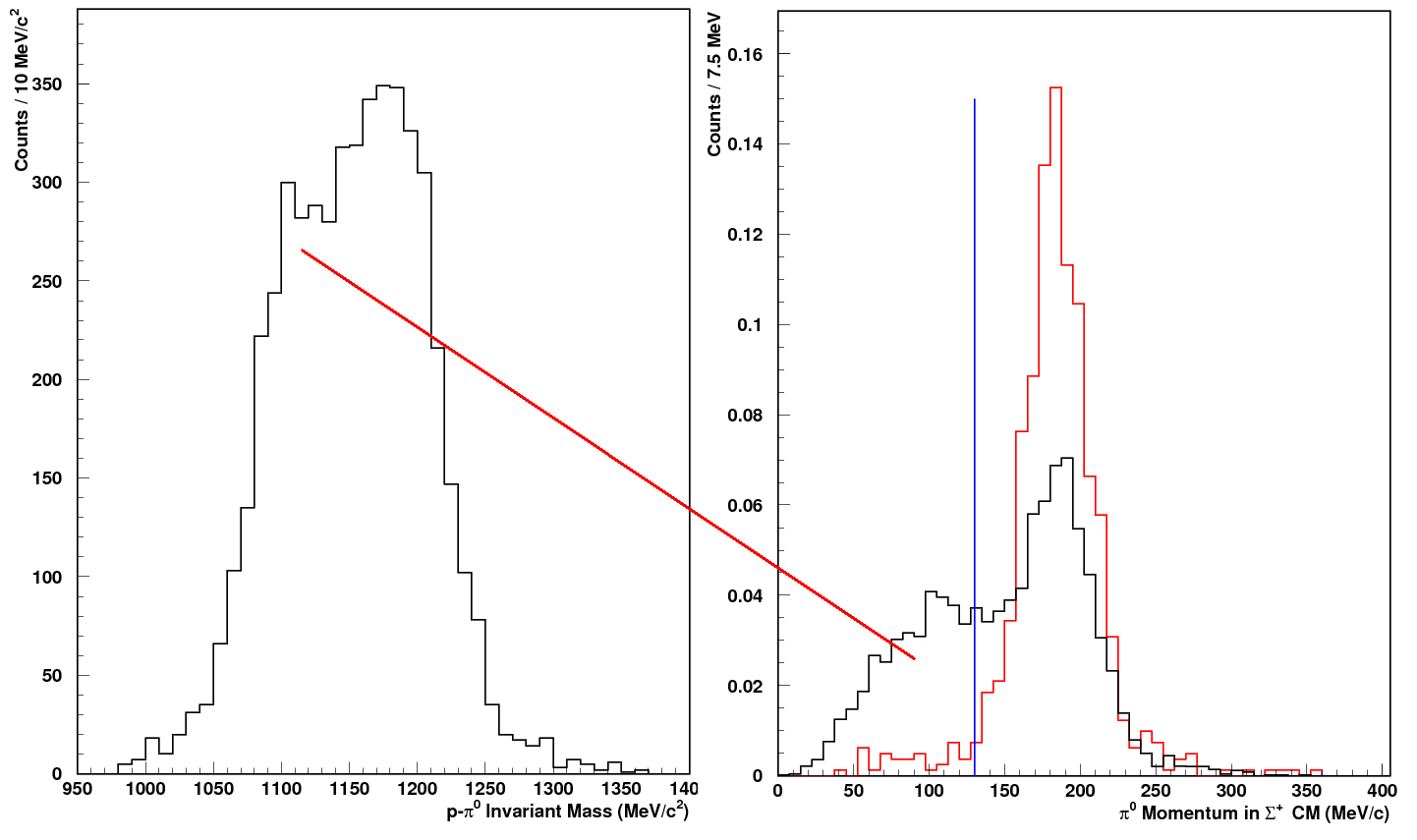}}
\caption{Left: $p\pi^0$ invariant mass spectrum; a clear double structure is seen with a peak at $\sim 1190\, MeV/c^2$ (the right $\Sigma^+$ mass value) and another one at lower masses.
Right: Data (black) and MC simulation (red) plots of the p (or $\pi^0$) momentum in the $\Sigma^+$ center of mass frame.}
\label{SHOULDERS}
\end{figure}

Once all the final particles of our events type are selected, the $p\pi^0$ invariant mass spectrum can be obtained to reconstruct the $\Sigma^+$;
the results are shown in the left part of fig. \ref{SHOULDERS} where a clear double structure is found.

\begin{figure}[htbp]
\centering
\mbox{\includegraphics[width=7.cm]{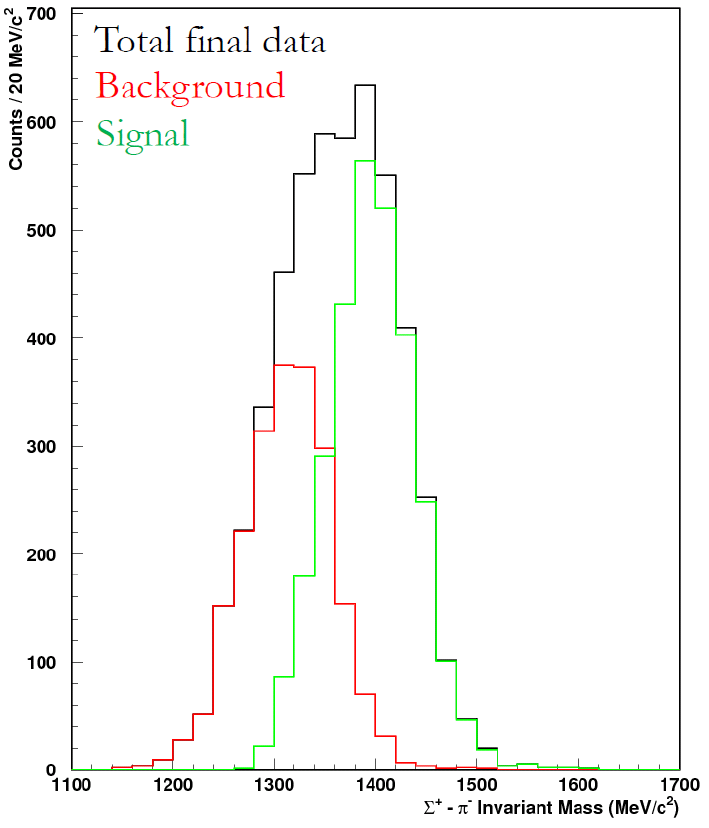}}
\caption{$p\pi^0$ total invariant mass spectrum (black) and individual spectra for the real $\Sigma^+$ (green) and for the low momentum component (red)}
\label{ASSOC}
\end{figure}

From the MC simulation, the two components can be separated when observing the proton (or $\pi^0$) momentum spectrum in the center of mass frame of the reconstructed $\Sigma^+$, as put in evidence in
in right part of fig. \ref{SHOULDERS}. When a real $\Sigma^+$ is correctly reconstructed, the CM momentum value has to be peaked around the nominal PDG value for the $\Sigma^+$ decay (189 MeV/c);
the red spectrum shows how, when a $\Sigma^+$ is simulated, the $\pi^0$ momentum is reconstructed at the right value. The lower momentum bump visible in the real data can be thus associated with 
the low mass shoulder in the $p\pi^0$ invariant mass. The association becomes evident when the $p\pi^0$ invariant mass is reconstructed individually for the two momentum components; resulting
spectra are shown in fig. \ref{ASSOC}

\subsection{Standard background analysis}

In order to explain the low mass peak, the possible sources of background have been studied. All the possible final states of a $K^-N$ interaction are listed in tab. \ref{STDBG}.

\begin{table}[htbp]
\begin{center}
\resizebox{.6\textwidth}{!}{%
\begin{tabular}{|l|c|c|c|r|}
\hline
Interaction	& Resulting particles	&  Final particles & B. R. & Status \\
\hline
$K^- p$  & $\Lambda\,\pi^0$   & $p\, \pi^-\,\pi^0$ &  64 \%  &  Cut on $\Lambda$ vertices \\
         &                    & $n\, \pi^0\,\pi^0$ &  36 \%  &  No $p\pi^-$ vertex \\
\hline
$K^- p$  & $\Sigma^+\,\pi^-$  & $p\, \pi^0\,\pi^-$ &  52 \%  &  SIGNAL \\
         &                    & $n\, \pi^+\,\pi^-$ &  48 \%  &  No $p\pi^-$ vertex \\
\hline
$K^- p$  & $\Sigma^0\,\pi^0$  & $\Lambda(p\pi^-)\,\gamma\, \pi^0 $ &  64 \%  &  Cut on $\Lambda$ vertices \\
         &                    & $\Lambda(n\pi^0)\,\gamma\, \pi^0 $ &  36 \%  &  No $p\pi^-$ vertex \\
\hline
$K^- p$  & $\Sigma^-\,\pi^+$  & $n\, \pi^-\,\pi^+$ &  100 \%  &  No $p\pi^-$ vertex \\
\hline
$K^- p$  & $\Lambda$   & $p\, \pi^-$ &  64 \%  &  No $p\pi^-$ vertex \\
         &             & $n\, \pi^0$ &  36 \%  &  No $p\pi^-$ vertex \\
\hline
$K^- p$  & $\Sigma^0$  & $\Lambda(p\pi^-)\,\gamma $ &  64 \%  &  Cut on $\Lambda$ vertices and no $\pi^0$ \\
         &             & $\Lambda(n\pi^0)\,\gamma $ &  36 \%  &  No $p\pi^-$ vertex \\
\hline
$K^- n$  & $\Lambda\,\pi^-$   & $p\, \pi^-\,\pi^-$ &  64 \%  &  Cut on $\Lambda$ vertices and no $\pi^0$ \\
         &                    & $n\, \pi^0\,\pi^-$ &  36 \%  &  No $p\pi^-$ vertex \\
\hline
$K^- n$  & $\Sigma^0\,\pi^-$  & $\Lambda(p\pi^-)\,\gamma\, \pi^- $ &  64 \%  &  Cut on $\Lambda$ vertices and no $\pi^0$ \\
         &                    & $\Lambda(n\pi^0)\,\gamma\, \pi^- $ &  36 \%  &  No $p\pi^-$ vertex \\
\hline
$K^- n$  & $\Sigma^-\,\pi^0$  & $n\, \pi^-\,\pi^0$ &  100 \%  &  No $p\pi^-$ vertex \\
\hline
$K^- n$  & $\Sigma^-$         & $n\, \pi^-$        &  100 \%  &  No $p\pi^-$ vertex \\
\hline

\end{tabular}}
\end{center}
\caption{Possible interactions between $K^-$ and nucleons}
\label{STDBG}
\end{table}

\begin{figure}[htbp]
\centering
\mbox{\includegraphics[width=6.5cm]{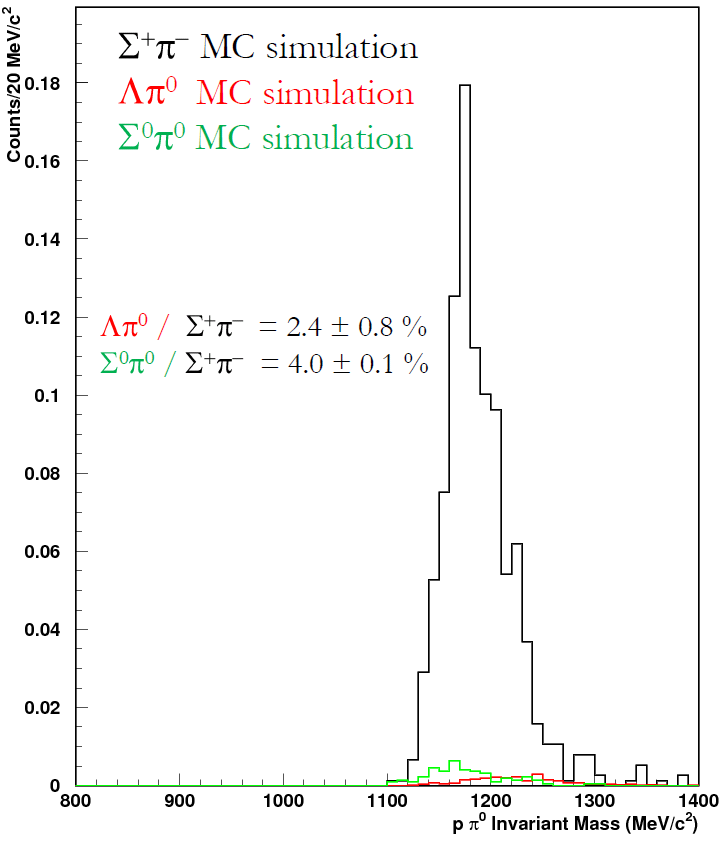}}
\caption{$\Lambda\pi^0$ (red) and $\Sigma^0\pi^0$ (green) MC simulations; the residual component ratio, whith respect to
$\Sigma^+\pi^-$ MC (black) are indicated in the spectra}
\label{MCBG}
\end{figure}

In the last column, an explanation of how these events are already excluded from the analysis is given. 
Where this rejection is made only via the cut on the $\Lambda(1116)$ described in the previous section,
detailed MC simulations have been done in order to understand how much, after the cut, a residual 
background component is still present. Results from this investigation are shown, for the accepted 
$\Sigma^+$ center of mass momentum region, in fig. \ref{MCBG}.

\noindent From the background ratios in fig. \ref{MCBG} it becomes evident that the standard hadronic background can not be the right 
explanation for the low mass peak in the $p\pi^0$ invariant mass spectrum.
These ratios are obtained with a normalization, in the number of events, between the MC of the background components and the signal spectra, taking into 
account the branching ratios available, unfortunately with big errors, in literature \cite{BR}.

\section{The internal conversion hypothesis}

\begin{figure}[htbp]
\centering
\mbox{\includegraphics[width=7.cm]{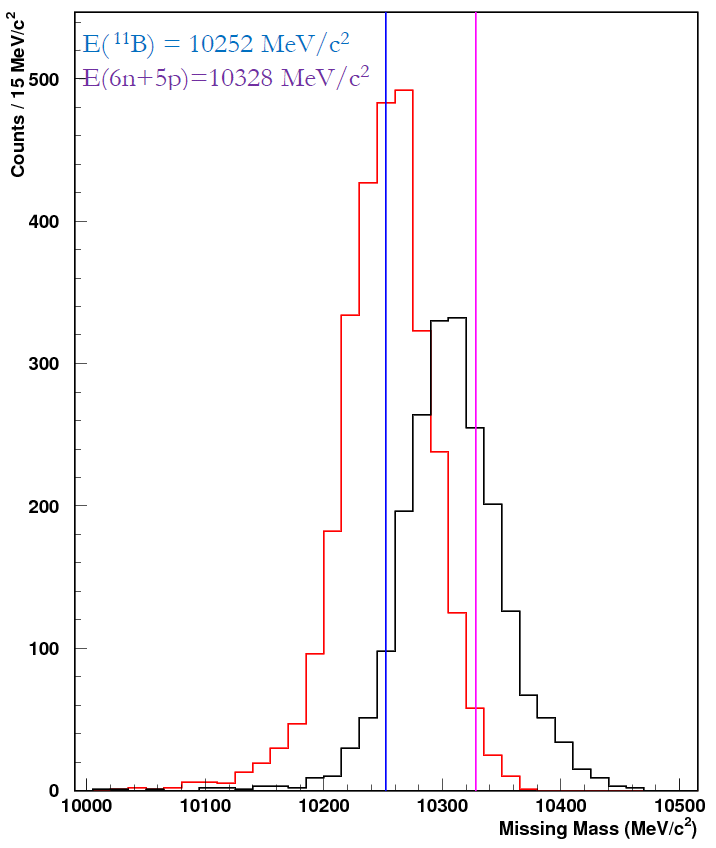}}
\caption{5n6p missing mass spectrum; the events corresponding to the real signal (selected from the $p\pi^0$ center of mass momentum) of the $\pi^0$ 
are shown in red, while the background events are shown in black. The blue line represents the $^{11}B$ mass while the violet one represents the free 5n+6p mass.}
\label{MISS}
\end{figure}

After the exclusion of the standard hadronic background, the remaining possibility is the internal conversion effect \cite{Internal}; 
this effect consists in an interaction of the $\Sigma^+$ with a nucleon of the same nucleus of the primary interaction, leading to the formation
of a $\Lambda(1116)$, which can then decay in a neutron and a $\pi^0$. Given the same rejection factors of tab. \ref{STDBG}, the possible remaining background
interactions are:

\begin{table}[htbp]
\begin{center}
\resizebox{0.5\textwidth}{!}{%
\begin{tabular}{|l|c|r|}
\hline
Interaction	& Int. Conv. Channel	&  Final particles  \\
\hline
$K^- p \rightarrow \Sigma^+\,\pi^-$   & $\Sigma^+\,+ \,n \rightarrow \Lambda\,p$   &  $n\,\pi^0\,p\,\pi^-$ \\
\hline
$K^- p \rightarrow \Sigma^+\,\pi^-$   & $\Sigma^+\,+ \,n \rightarrow \Sigma^0\,p$  &  $n\,\pi^0\,\gamma\,p\,\pi^-$ \\
\hline
$K^- p \rightarrow \Sigma^0\,\pi^0$   & $\Sigma^0\,+ \,n \rightarrow \Sigma^-\,p$  &  $n\,\pi^-\,p\,\pi^0 $ \\
\hline
\end{tabular}}
\end{center}
\caption{Possible final states after internal conversion in $K^-p$ interactions}
\label{ICBG}
\end{table}

\noindent with the second one involving also the presence of a third undetected photon.
According to these reactions, if internal conversion occurs the final particles are exactly the same as the final particles of the signal events;
the only difference are that the protons and the $\pi^0$ are not coming from a $\Sigma^+$ decay (leading to the low momentum component of fig. \ref{ASSOC}) 
and that a neutron is not bound in the spectator ($^{11}B$) of the primary interaction but is coming from the $\Lambda$ decay.
To test this hypothesis, a study of the missing mass of the remaining 5n6p, which should be different in the two cases was performed.
The results of this investigation are shown in fig. \ref{MISS}

According to fig. \ref{MISS}, as expected, the events corresponding to the signal events show a missing mass compatible with the $^{11}B$ one, which acts as spectator
in the primary interaction, while in the other events the only information that can be exctrated is that no complete fragmentation of the nucleus is observed, being the missing
mass peaked at lower masses than the free 5n+6p mass and indicating some residual binding energy between them.
This is today still a controversial subject and important information could be exctracted from further analysis.

\section{$\Sigma^+\pi^-$ invariant mass and momentum spectra}

\begin{figure}[htbp]
\centering
\mbox{\includegraphics[width=12.cm]{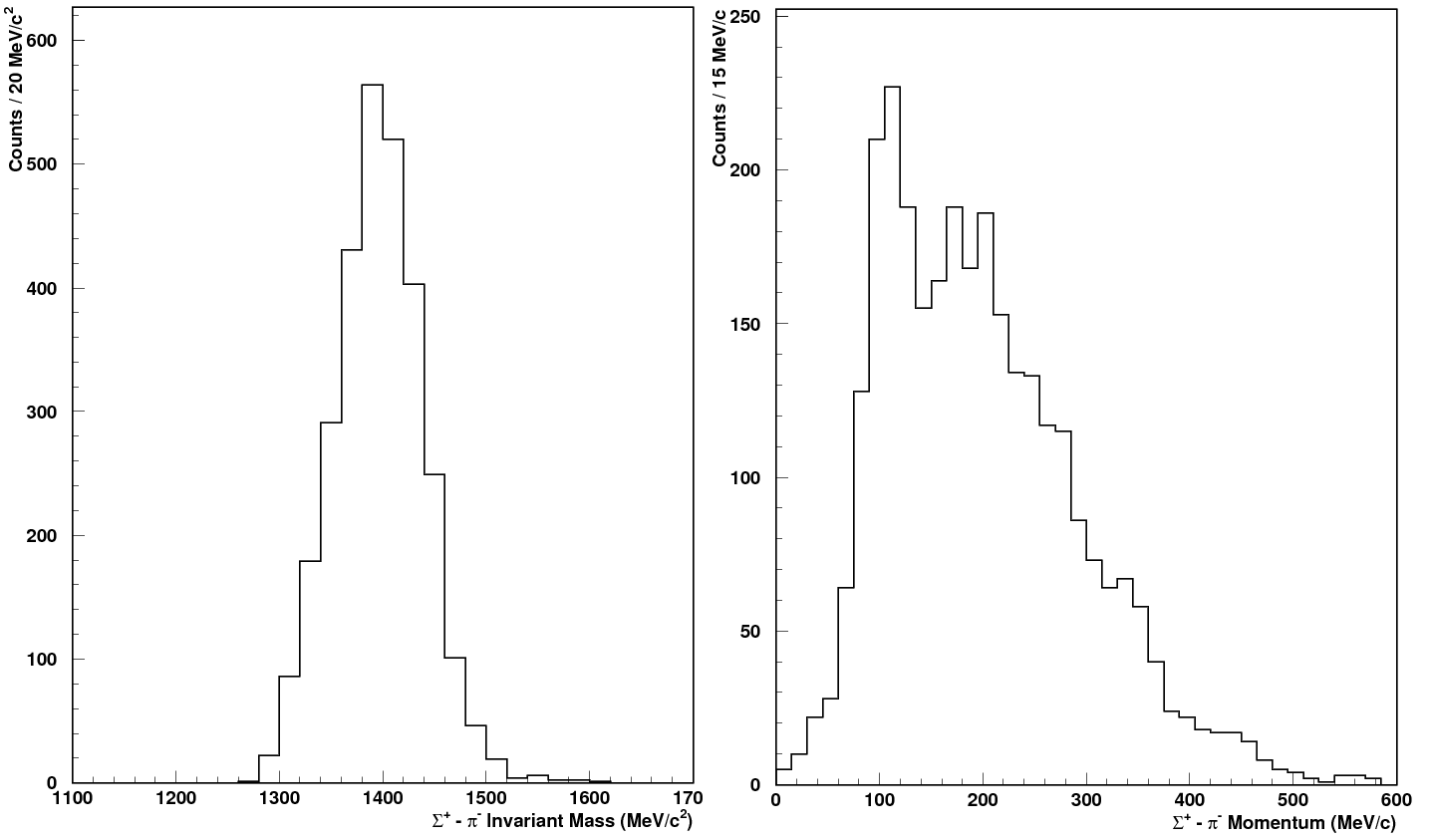}}
\caption{The final $\Sigma^+\pi^-$ invariant mass (left) and momentum (right) spectra.}
\label{FINAL}
\end{figure}

After the identification of the $\Sigma^+$ signal, the final $\Sigma^+\pi^-$ invariant mass and momentum spectra can be constructed. 
The plots are shown in fig. \ref{FINAL}.

It is important to notice that these final plots do not represents the "pure" $\Lambda(1405)$ resonance mass and momentum, since they are affected by the presence of the
$\Sigma(1385)$ resonance and of the non-resonant $K^-p$ interaction events, whose contribution has to be disentangled.
The momentum spectrum shows two components around 100 MeV/c and 200 MeV/c; this is in complete agreement with the results obtained in the analysis of
the $\Sigma^0\pi^0$ channel performed on the same data \cite{KRIS}, where these two components have been demonstrated to correspond respectively to in flight and at rest 
$K^-p$ interactions.

\section{Conclusions}

In this work we presented the preliminary results obtained for the study of the $\Sigma^+\pi^-$ channel in the $K^-p$ interaction in carbon 
with the KLOE detector at DA$\Phi$NE.
We discussed the obtained results considering the $\Sigma^+n$ internal conversion contribution, supporting this hypothesis with a missing mass spectrum of
the residual 5 neutrons and 6 protons. The study of this channel represents an open field with many questions to be answered, where important new information could be deduced from further analysis of our data. 
We showed also the $\Sigma^+\pi^-$ invariant mass and momentum spectra underlying their compatibility with the results obtained on the $\Sigma^0\pi^0$ 
events in the same data. 
A more detailed MC simulation of the three internal conversion channels events turned out to be necessary in order to quantify their contributions to the final spectra; 
in this way, it could be also possible to improve the existing value of the related branching ratios.
For what concerns the investigation of the $\Lambda(1405)$, an acceptance correction of the final spectra, together with a fit involving the two resonances and the non-resonant
events, is undergoing.

\section*{Acknowledgements}

We would like to thank Fabio Bossi, Stefano Miscetti, Erika De Lucia, Antonio
Di Domenico and Vincenzo Patera for the guidance and help in performing
the analyses. We thank as well Antonio De Santis for his support and
all the KLOE Collaboration for the fruitful collaboration.
Part of this work was supported
by the European Community-Research Infrastructure Integrating Activity ``Study of Strongly 
Interacting Matter'' (HadronPhysics2, Grant Agreement No. 227431, and HadronPhysics3 (HP3)
Contract No. 283286) under the Seventh Framework Programme of EU.


\begin{thebibliography}{99}

\bibitem{KLOE} F. Bossi, E. De Lucia, J. Lee-Franzini, S. Miscetti, M. Palutan and KLOE coll., \emph{Precision Kaon and Hadron Physics with KLOE}, Riv. Nuovo Cim. 31 (2008) 531-623
\bibitem{DAF} R. Baldini et al., \emph{Proposal for a Phi-Factory}, report LNF-90/031(R) (1990).
\bibitem{THEO1} J.C. Nacher, E. Oset, H. Toki, A. Ramos, \emph{Photoproduction of the $\Lambda(1405)$ on the proton and nuclei}, Physics Letters B455 (1999) 55-61.
\bibitem{THEO2} J. Esmaili, Y. Akaishi, T. Yamazaki, \emph{Experimental confirmation of the $\Lambda(1405)$ ansatz from resonant formation
of a K-p quasi-bound state in K-absorption by $^3He$ and $^4He$.}, Physics Letters B 686 (2010) 23-28.
\bibitem{CLAS} K. Moriya et al. (CLAS Collaboration), \emph{Measurement of the $\Sigma \pi$ Photoproduction Line Shapes Near the $\Lambda(1405)$}, Phys. Rev. C 87, 035206 (2013).
\bibitem{KRIS} K. Piscicchia, \emph{$\Lambda(1405)$ measurement through the decay to $\Sigma^0\pi^0$, resulting from $K^-$ absorption on $^4He$ and $^{12}C$, with the KLOE detector}, Ph.D. thesys (2013) and contribution in the present volume
\bibitem{Internal} C. Vander Velde-Wilquet et al., \emph{The conversion probability and the emission ratio of charged $\Sigma$-hyperons following $K^-$ meson capture at rest in carbon}, Nucl. Phys. A241 (1975) 511.
\bibitem{DriftChamber} M. Adinolfi et al., \emph{The tracking detector of the KLOE experiment}, Nucl. Inst. Meth. A 488, (2002) 51.
\bibitem{Calo} M. Adinolfi et al., \emph{The KLOE electromagnetic calorimeter}, Nucl. Inst. Meth. A 482, (2002) 364.
\bibitem{PDG} J. Beringer et al. (Particle Data Group), \emph{The review of particle pyhisics}, Phys. Rev. D86, 010001 (2012)
\bibitem{BR} C. Vander Velde-Wilquet et al., \emph{Determination of the Branching Fractions for $K^-$-Meson Absorptions at Rest in Carbon Nuclei},  Nuovo Cimento 39 A, (1977) 538.

\end{thebibliography}
\end{document}